\documentclass{article}

\usepackage{microtype}
\usepackage{graphicx}
\usepackage{subfigure}
\usepackage{booktabs} 
\usepackage{color,soul} 
\usepackage{todonotes} 
 
\usepackage{hyperref}

\usepackage[preprint]{neurips_2021}

\usepackage{amsmath}
\usepackage{amssymb}
\usepackage{mathtools}
\usepackage{amsthm}

\usepackage[capitalize,noabbrev]{cleveref}

\theoremstyle{plain}

\theoremstyle{definition}

\theoremstyle{remark}

\title{The Response Shift Paradigm to Quantify Human Trust in AI Recommendations}

\author{
  Ali Shafti\\
  Brain \& Behaviour Lab\\
  Department of Computing\\
  Imperial College London\\
  London, UK, SW7 2AZ \\
   \And
   Victoria Derks \\
   Brain \& Behaviour Lab\\
   Department of Computing \\
   Imperial College London \\
   London, UK, SW7 2AZ \\
   \And
   Hannah Kay \\
   Brain \& Behaviour Lab\\
   Department of Computing \\
   Imperial College London \\
   London, UK, SW7 2AZ \\
   \And
   A. Aldo Faisal\thanks{Behaviour Analytics Lab, Data Science Institute, SW7 2AZ, London, UK; UKRI CDT in AI for Healthcare, Imperial College London, SW7 2AZ, London, UK; MRC London Institute of Medical Sciences, W12 0NN, London, UK.}\\
   Brain \& Behaviour Lab\\
   Department of Computing \& Bioengineering \\
   Imperial College London \\
   London, UK, SW7 2AZ \\
   \texttt{aldo.faisal@imperial.ac.uk} \\
}

\begin{document}
\maketitle

\begin{abstract}
Explainability, interpretability and how much they affect human trust in AI systems are ultimately problems of human cognition as much as machine learning, yet the effectiveness of AI recommendations and the trust afforded by end-users are typically not evaluated quantitatively. We developed and validated a general purpose Human-AI interaction paradigm which quantifies the impact of AI recommendations on human decisions. In our paradigm we confronted human users with quantitative prediction tasks: asking them for a first response, before confronting them with an AI's recommendations (and explanation), and then asking the human user to provide an updated final response. The difference between final and first responses constitutes the shift or sway in the human decision which we use as metric of the AI's recommendation impact on the human, representing the trust they place on the AI. We evaluated this paradigm on hundreds of users through Amazon Mechanical Turk using a multi-branched experiment confronting users with good/poor AI systems that had good, poor or no explainability. Our proof-of-principle paradigm allows one to quantitatively compare the rapidly growing set of XAI/IAI approaches in terms of their effect on the end-user and opens up the possibility of (machine) learning trust.

\end{abstract}

\section{Introduction}
\label{introduction}
Human-facing Artificial Intelligence (AI) is making increasing progress from toy scenarios to real-world deployment with safety-critical applications e.g. in healthcare or aerospace industries -- little AI will ever be used outside a scenario where it interacts with humans directly.
The recent movement towards Explainable and Interpretable Artificial Intelligence \citep[e.g.][]{doshi2017towards,rudin2019stop} aims to account for this role of AI, and to create AI models where the AI's output and decision making process can be better understood by its human users, whilst maintaining high predictive performance.

Building explainability and interpretability into AI recommender systems is a first step towards thinking about how to engineer trust into Human-AI interactions. A trusted AI system will be used and adopted (taken into long-term use), its recommendations followed and not ignored as often seen with ``smart'' alerts in hospital settings \citep{bedoya2019minimal}. More importantly a trusted AI decision support system can be incorporated into human decision making workflows as a member of the team -- by allowing the AI to weigh in on the human decision \citep{gille2020we}. Yet, much of the XAI/IAI development history has been focused on algorithms that were thought up so as to give better explanations, without a measure of the effect they might have on trust. 

IAI is particularly essential when dealing with human-facing AI, and in safety conscious environments, e.g. medicine \citep{wiens2019no}, where the ultimate decisions on diagnosis and treatment will be in the hands of medical experts, even when they use AI models to aid these decisions \citep{komorowski2018artificial}. Consequently, whether interacting humans trust the output of AI models becomes an important factor in their usefulness, hence the need for interpretability \citep{payrovnaziri2020explainable}.

To enable AI deployment in regulated settings with advanced application domains, as in healthcare, co-creation teams of human factors experts, clinical experts and AI researchers have started to design and evaluate the usefulness of an AI system's human interface -- yet often this happens after much of the machine learning research has completed. We as machine learning researchers can come up with strategies for explainability or interpretability, but ultimately it is the human users themselves who are the true arbiters of trust, the quality of explanations and the usefulness of interpretations. Crucially, the explanation and interpretation development process is not automatable nor particularly quantitative. Eventually, we want to be able to machine learn how to improve explanations, interpretations and steer trust components. Towards this we present here a quantitative assessment framework which can be applied from the start of the machine learning development process. 

Here we propose a simple way of measuring and quantifying trust in AI systems in the form of a Human-in-the-Loop protocol. Our approach is inspired by Bayesian Decision theory as successfully applied by cognitive sciences and neuroscience \citep{tenenbaum2006theory,kording2006bayesian} to model and predict human decision making behaviour quantiatvely. We directly assess how much an AI can shift actual human decisions and responses, under different controlled conditions. It is a general-purpose protocol that can be systematically used in supervised learning, specifically regression settings. Our approach can deal with the natural variability of human decision making and actions \citep{faisal2008noise}, extracts more information per interaction by using a continuous report (instead of categorical or ordinal questions) and in general avoids relying on subjective self-report or momentary assessments (``Rate how much you trust...'') or choice of automation level. We present in the following, the model for an example AI recommendation task based on tabular data (predicting the school grade of pupils), derived explanations, the Human-in-the-Loop protocol and the experimental results from running the experiment on hundreds of users online.

\section{Related Work}
Trust is a latent construct in a human's mind which is challenging to measure explicitly \citep{toreini2020relationship}. Previous work used self-report measures such as a subjective rating with questionnaires or rating scales \citep{ajenaghughrure2020measuring} asking users to indicate their degree of trust; however, self-report measures are too subjective to be viable in applied settings and are difficult to use due to their inherent variability for e.g. automating trust engineering which may require determining gradients of trust. Quantitative 
physiological and neural measures have been proposed, which use social signal cues such as gaze behaviour \citep[e.g.][]{walker2019gaze,lu2019eye} or physiological measures such as EEG and GSR \citep[e.g.][]{akash2018classification,wang2018eeg,gupta2020measuring}. Although these measures could be used for dynamic tracking  they  require special hardware, and might suffer from ambiguities and confounders due to the indirect measurement and in many use cases cannot be used for life-long calibration of trust.
In contrast, behavioural measures where the user's actual actions and decisions are measured, are useful for applications in real-world situations. Examples of behaviour used as a trust measure includes applications in choosing manual or automatic tasks \citep[e.g.][]{okamura2020adaptive} or choosing an automation level \citep[e.g.][]{drnec2016trust, riley2018operator}. Behaviours might not be observable if there is no interaction; nevertheless, behavioural measures are practical and can easily be used as a basis for modeling and prediction.

Previous studies have shown that humans do not behave rationally when interacting with machine learning algorithms and will often default to trusting in human judgement over a machine \citep{dietvorst2015algorithm}. Trust is therefore an important consideration when creating explainable AI models. Self-report measures such as Likert scales are a common way to measure trust \citep{buccinca2021trust,bansal2021does,alarcon2021exploring}. Other experiments take trust as a binary value, where a participant is said to trust the model only if they follow the recommendation of the model \citep{lai2019human}. Self-report measures as a measure of trust are especially problematic, as they are well known to be affected by biases such as the perceived desirability of answering in a certain manner and are inconsistent between subjects, over-relying on the subject’s perception of themselves \citep{paulhus2007self}. Trust in AI models has been studied previously for classification tasks such as disease diagnosis \citep{anton2022trust,alam2021examining}. This is, however, not applicable to AI models that make predictions on a continuous scale. For example, the AI Clinician \citep{komorowski2018artificial} makes recommendations about the precise amounts of IV and vasopressor that should be delivered to ICU patients. When a human is faced with a situation where their own prediction or recommendation is different from that of an AI model, there are an infinite range of potential choices in between the two recommendations which they might select.

Trust in AI as an abstract concept is relational, highly complex and involves at least two actors: one actor trusts the other actor to act in the right way. This relationship is influenced by diverse framing factors — culture, belief systems and context among others \citep{vollmer2018machine}. Here we dramatically simplify this complex concept of trust into a measurable quantity through the shift in a human's response after being exposed to the AI's recommendation. This allows us to: A. put the development of trust in AI, which is ultimately a problem situated in the human domain, into the domain of sound scientific evidence, and B. to consider AI methods that can start learning from or improve trust in Human-AI interactions by using a trust measure as a signal for machine learning \citep[e.g.][]{okamura2020adaptive}.

Our approach is inspired by Bayesian decision theory and cue combination experiments \citep{yuille1996bayesian}. Experimental psychology studies have shown that humans use Bayesian inference to combine cues from multimodal sources \citep{beierholm2008comparing,jones2016tutorial}. While this is often studied in the context of sensory modalities \citep{kording2006bayesian,maloney2009bayesian}, it can also be used to interpret how people combine their own beliefs with an external recommendation to make a decision. Our framework thus allows us to objectively quantify human trust in an AI model's recommendation.

\section{Methods}
Our study aims to investigate human trust in AI which makes predictions on a continuous scale and how different factors might affect this trust, including the quality of the AI, inclusion of explainability and the quality of said explainability, among other factors such as demographics and opinions on AI. We choose a grade prediction task, where given tabular data about student backgrounds, participants (and our AI) need to predict the grade of the student. Our AI predictions and the relevant explainability come from real trained models.

\textbf{Grade prediction task -- }The tabular data task uses the publicly-available Student Performance Data Set \citep{cortez2008using}. This dataset contains the grades of Portuguese students on three separate high school tests, as well as additional information about the students such as their number of absences and the jobs of their parents. See Appendix~\ref{appendix:tabular_features} for the full list. The task for participants in the experiment is to predict a student’s grade on their final test from the other information about the student. As grades on the first two tests are highly correlated with performance on the final test, these two initial tests are excluded from the features. 

\begin{figure}[ht]
    \begin{center}
    \includegraphics[width=.8\textwidth]{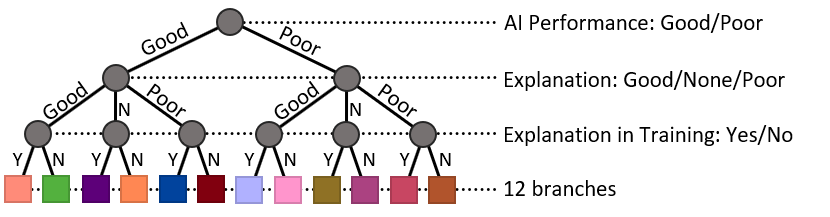}
    \caption{Our experiment consists of 12 branches, based on conditions involving the quality of the AI, the quality of explanations, and the inclusion or exclusion of explanations during training and testing phases (see Fig.~\ref{fig:protocol}).}
    \label{fig:branches}
    \end{center}
\end{figure}

\textbf{AI prediction manipulation -- }We used a simple feedforward neural network as our grade prediction AI. The model consisted of an input layer with 43 neurons, two hidden layers, one with 32 and the next with 16 neurons, and an output layer with one neuron, using ReLU activations except for the output neuron. Categorical features were dummy encoded, and no normalisation or scaling was performed for numerical features or labels. The model was implemented in PyTorch \citep{paszke2019pytorch}, trained for 100 epochs with an MSE loss, using Adam optimiser \citep{kingma2014adam}, and a batch size of 5. 

For our factorial experiment design (see Fig.~\ref{fig:branches}), we needed two grade prediction AIs of different performance levels, to represent our Good AI and Poor AI conditions. Nonetheless, the output of the Poor AI still needed to look reasonable to human participants. The two AIs were obtained by training them using different learning rates. For the Good AI, a learning rate of $0.003$ was ideal and led to an RMSE of $2.90$. For the Poor AI, a learning rate of $0.00065$ led to an RMSE of $4.1$. This could be worsened with other learning rates but that would lead to implausible grades. 

\textbf{AI explainability manipulation -- }To assess the effect of explainability on AI-assisted human responses, we produce explanations for our AI, using LIME \citep{ribeiro2016should}, a well-established method and one of the most popular (Average $1450$ citations per year since publication according to Google Scholar). LIME aims to produce interpretable post-hoc explanations, by using simple machine learning models such as linear regression or decision trees. For tabular datasets like ours, an input sample is perturbed by picking different values for numerical and categorical features, and swapping the values for binary features. The explanation then takes the form of a figure that demonstrates the features which contribute most positively and negatively to the prediction. Positive features are those features which lead to an increase in the predicted grade if they are included. Negative features lead to a decrease in the grade, with the final output being a balance of positive and negative features. Weights of the features show how much the features contributed and therefore which features contributed most. See screenshots of our experiment interface in Appendix~\ref{appendix:experiment_interface}, for an example of the LIME explanation visualisations.

For the factorial design in our experiment, we require two explanations, serving as Good and Poor Explanations. LIME currently does not have a metric to assess the quality of its explainability output. We decided to cross-use the explanations to make up for this, i.e. an explanation is considered a good explanation, if it corresponds to the AI model you are facing. Conversely, a poor explanation will be one that does not correspond to your AI model. As an example, if a participant is facing the Poor AI, and the explanation meant for the Good AI, then we will consider them as having faced a Poor Explanation.

\begin{figure}[t]
    \begin{center}
    \includegraphics[width=0.7\textwidth]{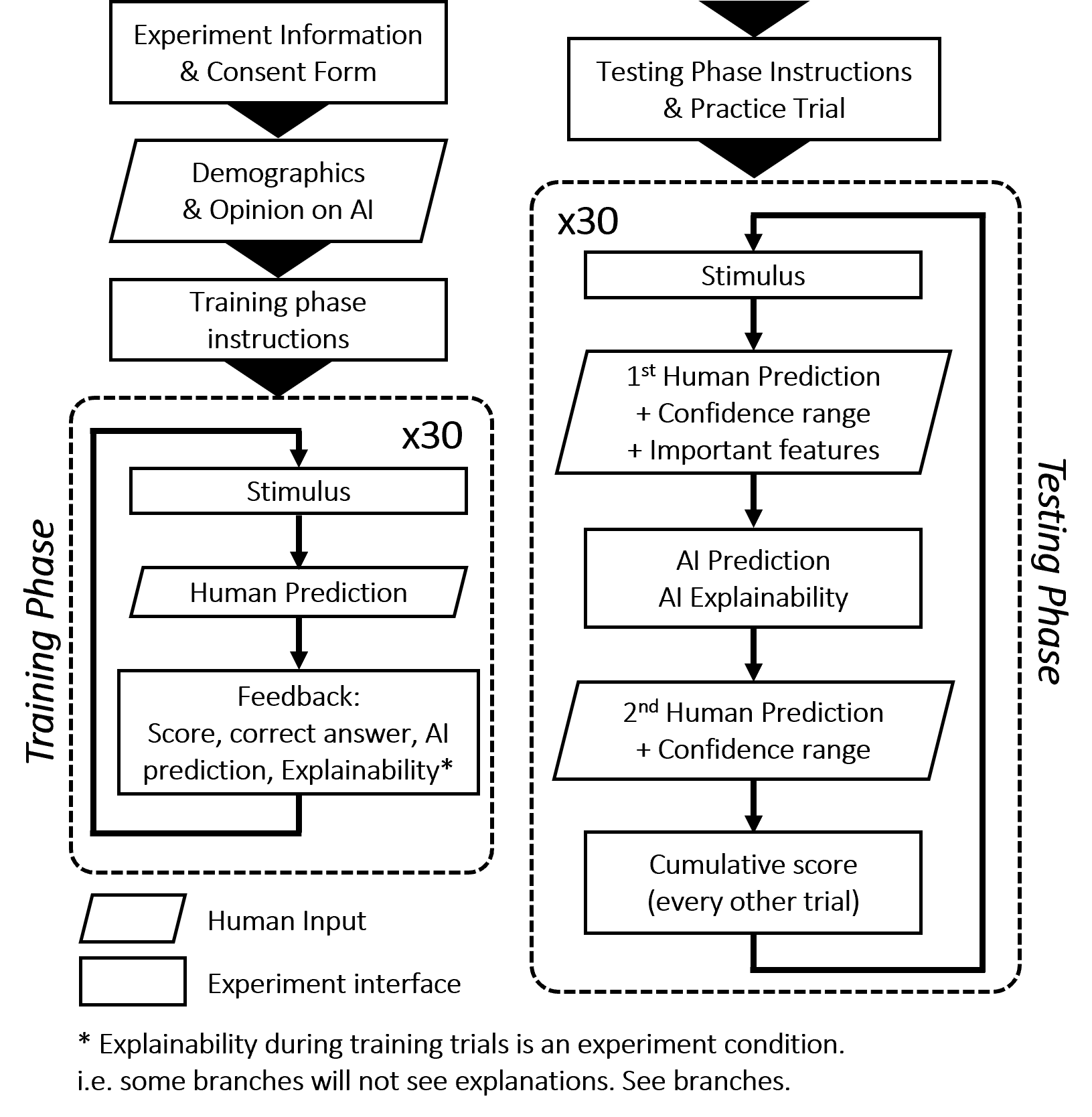}
    \caption{Flow chart of our training and testing phases for our Human-in-the-Loop protocol.}
    \label{fig:protocol}
    \end{center}
\end{figure}

\textbf{Crowd sourcing -- }Amazon Mechanical Turk (often referred to as MTurk) is a crowdsourcing platform where `requesters' (experimenters) can request completion of tasks from `workers' (participants). Each task completion earns the worker a reward specified by the requester. Additionally, requesters can provide workers with a bonus reward, which can be based on their performance during the task. This system of payment works well for our study, as participants can be rewarded without directly collecting personal details like banking information.

Our experiment was created as a webpage using HTML and JavaScript. The jsPsych \citep{deLeeuwJR2015JsPsych:Browser} library was used for implementation, as it provides a framework for conducting online psychological style experiments. The main use of this library was to aid with the creation of multiple trials of a particular type, and to measure the reaction time of participants for each trial. The library offered a plugin template, which was used to create several specific plugins for each part of a trial. In addition, there were several plugins which were used for gathering demographic information and displaying instructions to participants.

The webpage was hosted on our lab server. This was implemented using Docker and node.js. The jsPsych library returns a JSON object containing all of the experimental data collected during the experiment. Participant data was stored in a SQLite database on this server, only if participants completed the entire experiment. 

\textbf{Human-in-the-Loop protocol -- }The experimental procedure overview is shown in Fig.~\ref{fig:protocol}. Participants start on an overview page that shows them the sections to follow, and roughly how much time each section is expected to take. Once they enter the experiment, they are asked to read an information sheet and to provide their consent to take part in the study. Our study has received ethical approval from the Imperial College London Science Engineering Technology Research Ethics Committee, reference number 20IC6224. Participants’ explicit consent is required for participation in the study, but they can opt out of their data being used in different supporting studies. After consenting, participants are asked demographic questions regarding their gender, age and education level. In addition, participants answer three questions regarding their attitude towards AI, using a 5-point Likert scale (see Appendix~\ref{appendix:questionnaire} for a list of questions). Once this part of the experiment is complete, participants are given further instructions regarding the task they are about to perform. 

Our experiment involves two phases, training and testing, each preceded by their own instructions page. The aim of the training phase is to give the participants a chance to get to know the grade prediction task, but also to familiarise them with the AI and explainability they will be facing, so that they form a prior on the AI's performance. To this end, in the training phase, participants are faced with an stimulus (i.e. student features, as listed in Appendix~\ref{appendix:tabular_features}), and asked to provide their grade prediction. Once the prediction is submitted, the participants are provided with feedback: the score they obtained, what the correct answer was, what the AI predicted, and (if they are in an explainability branch) what the explanation for the AI prediction is. This is repeated for 30 trials, see Fig.~\ref{fig:protocol}.

The training phase is followed by the testing phase. The participants are provided with a stimulus and are asked to input their grade prediction response, mark the features they used to make their decision (this allows us to consider their overlap with the AI, based on LIME output), and provide a confidence range for their prediction (see Fig.~\ref{fig:interface} and Fig.~\ref{fig:further-interface} in Appendix~\ref{appendix:experiment_interface}). The participants will see an estimate of what their score would be if they were correct as they set the confidence range; the larger the range, the lower the score they will receive which means a lower cash bonus prize at the end of the experiment (see Appendix~\ref{appendix:experiment_interface} for screenshots of the interface). Once participants submit this, they are presented with the AI's prediction for the same stimulus, and (if they are in an explainability branch) the explanation for that prediction. At this point, the participants are given a chance to update their response. Once this is submitted, at every other trial, participants are provided with their cumulative score for motivation purposes. This process starts with a practice trial, followed by 30 actual trials, see Fig.~\ref{fig:protocol}. At the completion of the testing phase, the participants are given the chance to provide any feedback, and provided instructions on how to receive their payment through MTurk.

The above experimental procedure is implemented in 12 branches. These branches are based on combinations of the following conditions: Interacting with a Good AI vs. a Poor AI; seeing Good, Poor or No explanations in the testing phase; and receiving or not receiving explanations in the training phase (the type of explanations follows that of the testing phase, with a good explanation used for cases where there is no explanation during testing). See Fig.~\ref{fig:branches} for a breakdown of the branches. This design allows us to control for different conditions and see the effects of AI performance and explanations on human responses after AI input.

\begin{figure}[ht]
    \begin{center}
    \includegraphics[width=\textwidth]{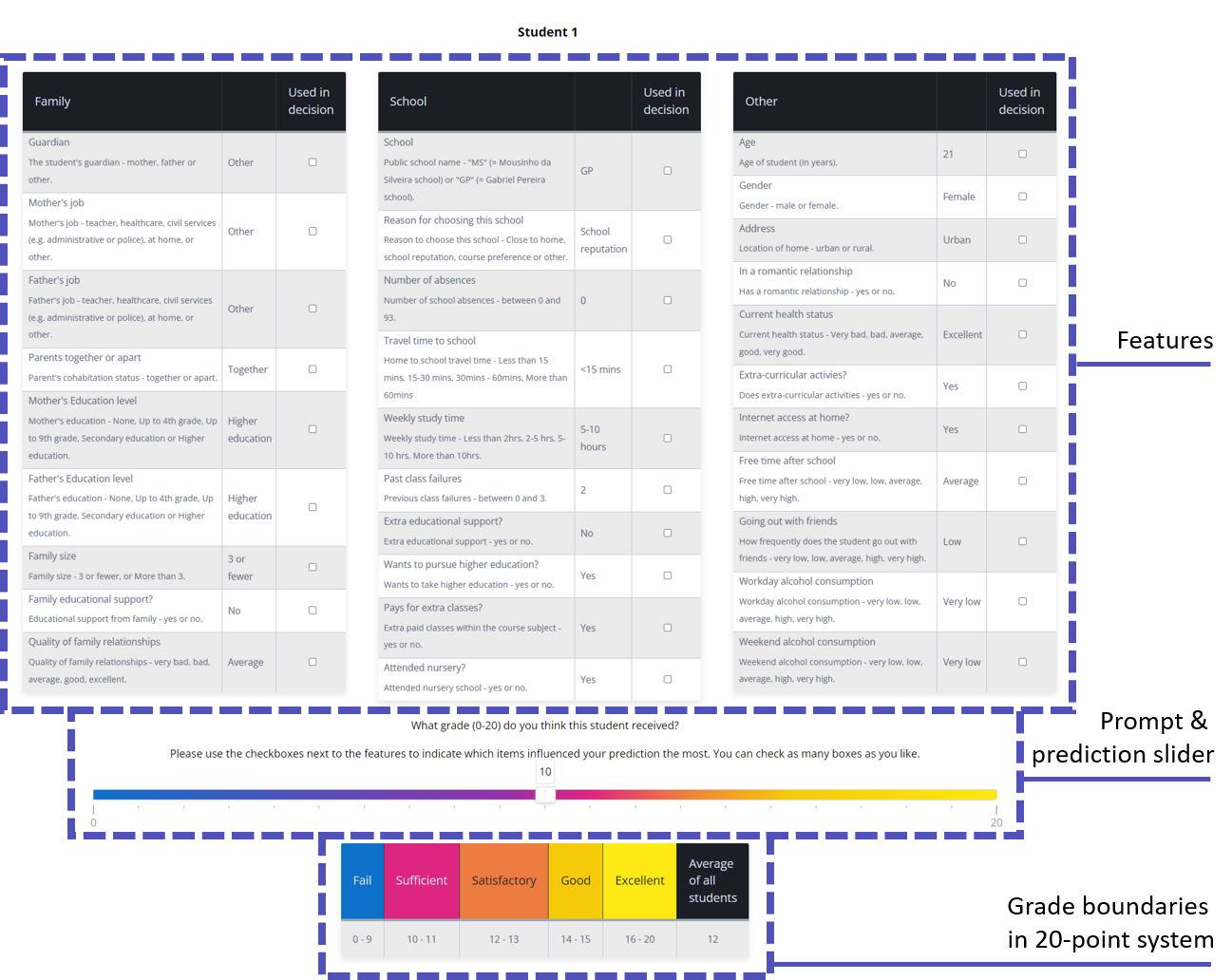}
    \caption{Web-interface for human subjects, accessed through Amazon Mechanical Turk. Top area shows the individual student's data with a tick box per feature for subjects to tick subjectively important features. Middle area contains a slider to indicate the predicted grade. Bottom area contains a reminder of the grading scheme (colour coded with the slider, and giving grade average). AI recommendations and explanations are displayed in a similar interface, with the slider to input the 2nd response (see Fig.~\ref{fig:further-interface} in Appendix~\ref{appendix:experiment_interface}).}
    \label{fig:interface}
    \end{center}
\end{figure}

\section{Results}

Our experiments are continuously running on MTurk. At the moment, we have obtained results from 167 participants. The participants are $27.5\% / 72.5\%$ female/male\footnote{Choices were: Female, Male, Non-binary, Other, Prefer not to say -- but so far only Male/Female choices were made}, and aged $36.4\pm10.1$ (mean $\pm$ standard deviation). Demographics can be seen in Fig.~\ref{fig:demographics}, in Appendix~\ref{appendix:demographics}.

\begin{figure}[h]
    \begin{center}
    \includegraphics[width=0.5\textwidth]{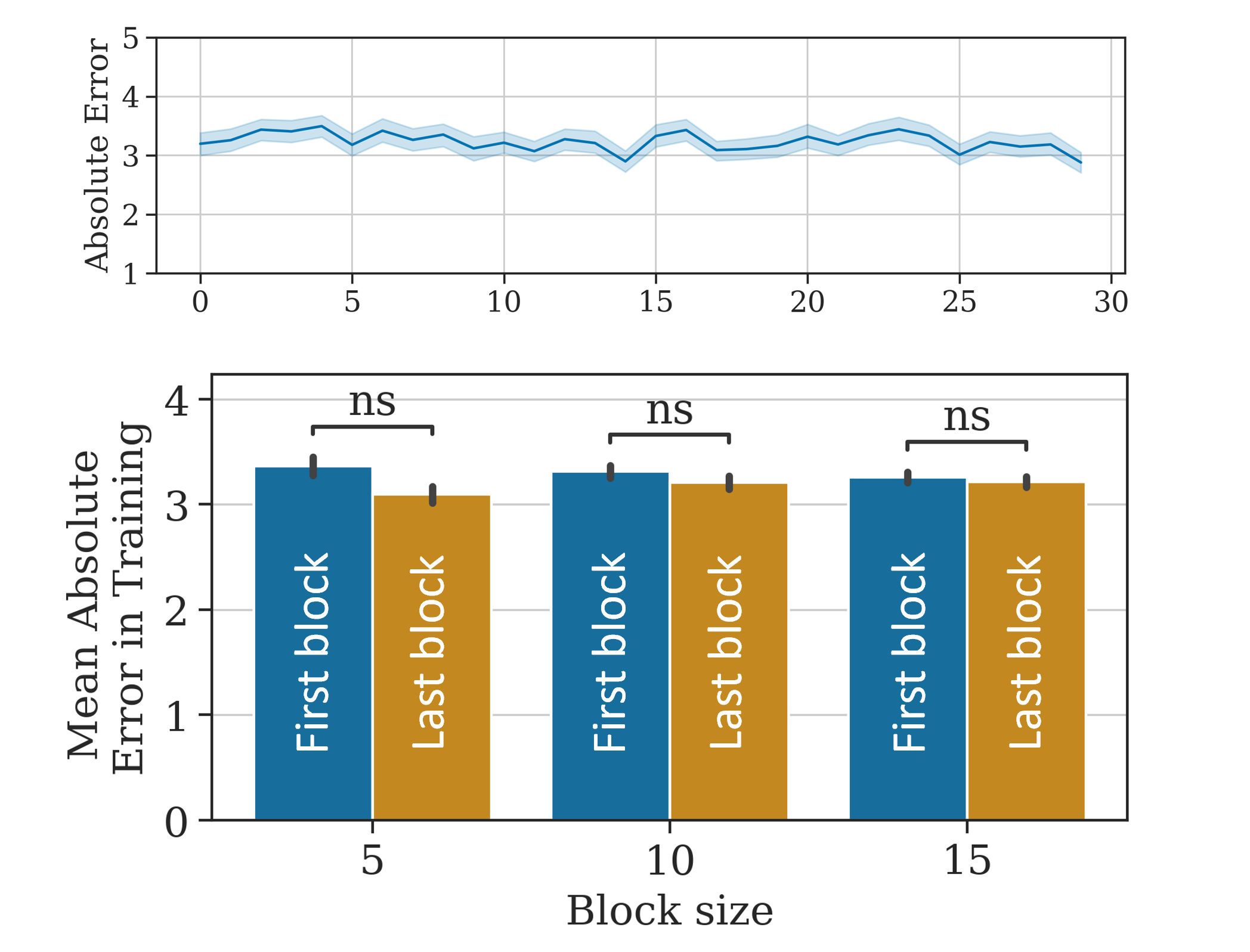}
    \caption{Performance of participants in the training phase. Top: Absolute error throughout the trials, showing mean and standard error, with no learning effect. Bottom: Comparing performance levels at the beginning and end of the phase, by breaking the trials down into blocks of size 5, 10 and 15, and comparing mean absolute error. We see no significant differences, indicating no learning.
    \label{fig:training}}
    \end{center}
\end{figure}

Our training phase is designed to familiarise the participants with the AI and potential explanation they will be facing. Additionally, the training phase is there for the participants to get to know the prediction task, so that we remove any learning effects from our actual AI interaction trials. Figure~\ref{fig:training}, top, shows the absolute error results of our participants during the training phase. We see no learning effect, with a steady performance throughout, further validated by breaking down the trials into blocks of sizes 5, 10 and 15, and then comparing the first and last blocks for significant differences in performance. We see no signifcant differences in any of the block size cases (two-sided Mann-Whitney-Wilcoxon test, with Bonferroni correction), see Figure~\ref{fig:training}, bottom. 

\begin{figure}[b]
    \begin{center}
    \includegraphics[width=\textwidth]{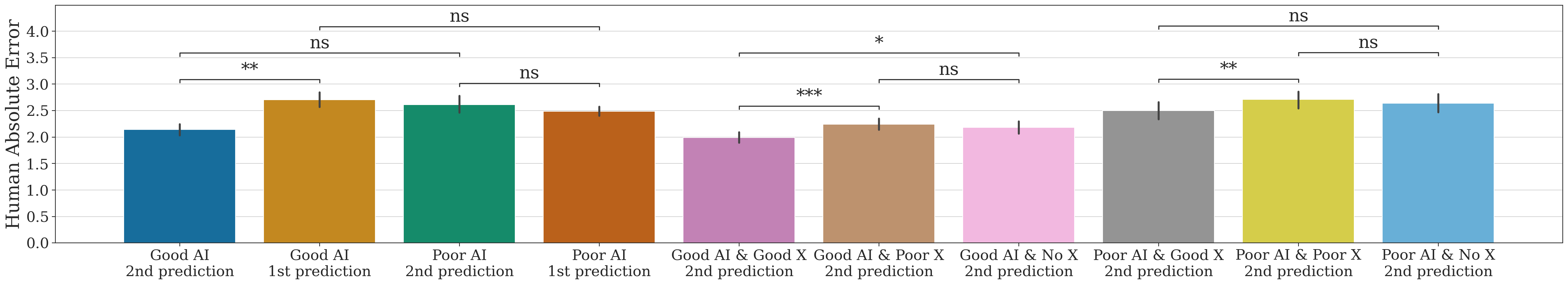}
    \caption{Prediction error of the human subjects under different AI interaction conditions. The error is measured as the difference between the actual value and the predicted value. The columns are the errors of human users predicting the answer while being exposed to different branches of our explainability paradigm (see main text for details): Good/Poor AI=AI prediction quality seen by user, Good/Poor X=AI explanation quality seen by user, No X=no explanation seen. Horizontal square brackets indicate test for significant statistical differences between different predictions errors   (ns = not significant, * = 5\% level, ** = 1\% level). Bars are mean $\pm$ standard error.
    \label{fig:error}}
    \end{center}
\end{figure}

For the purposes of this work, we look into the main questions regarding the effect of AI performance and the quality of explanations on the influence of AI over human decisions, and the subsequent performance of humans during the testing phase of our protocol. Fig.~\ref{fig:error}, inspects the role of AI performance and explanations, on human performance, considering mean performances per stimuli across subjects, and then comparing the grand average across stimuli. We see that humans paired with the Good AI reduce their error significantly (paired t-test, with Bonferroni correction, $p<0.01$) on their updated response (i.e. after they receive the AI input). This reduction is not significant when humans are paired with a Poor AI, suggesting that the Poor AI's recommendation is not helping the human decision maker. 

When breaking the results down by the explainability conditions, and looking only at the updated response errors (i.e. final performance, after exposure to the AI prediction), we see that with the Good AI, having a Good Explanation leads to significantly lower error, when compared with the case of having a Poor Explanation ($p<0.01$), or No Explanation ($p<0.05$, see Fig.~\ref{fig:error}). However, there is no significant difference between having Poor Explanation or No Explanation, suggesting that a poor explanation will not be helpful, to the point that outcomes will be similar without explanation.

\begin{figure}[t]
    \begin{center}
    \includegraphics[width=\textwidth]{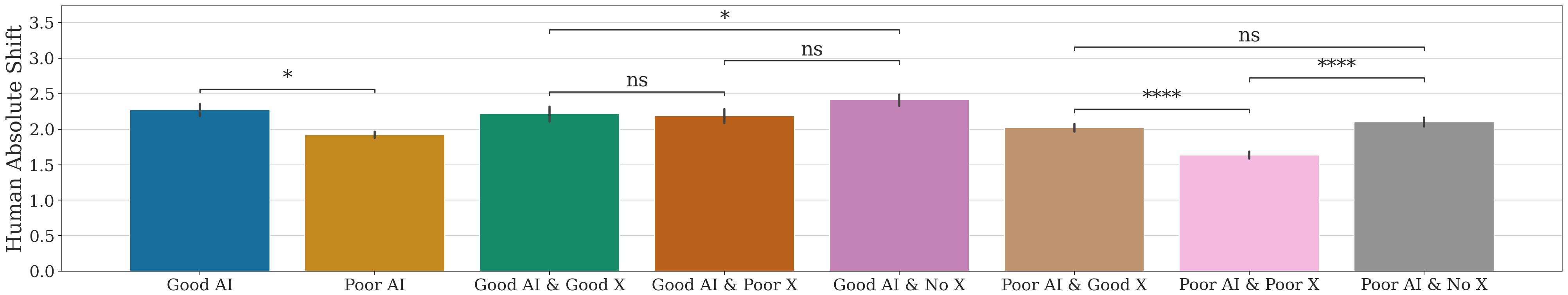}
    \caption{Shift of human response between original estimate and revised estimate after having been exposed to the AI. The columns are the absolute response shift of human users while being exposed to different branches of our explainability paradigm (see main text for details): Good/Poor AI=AI prediction quality seen by user, Good/Poor X=AI explanation quality seen by user, No X=no explanation seen. Horizontal square brackets indicate test for significant statistical differences between different predictions errors (ns = not significant, * = 5\% level, **** = 0.01\% level). Bars are mean $\pm$ standard error.}
    \label{fig:shift}
    \end{center}
\end{figure}

\begin{figure}[b]
    \begin{center}
    \includegraphics[width=0.5\textwidth]{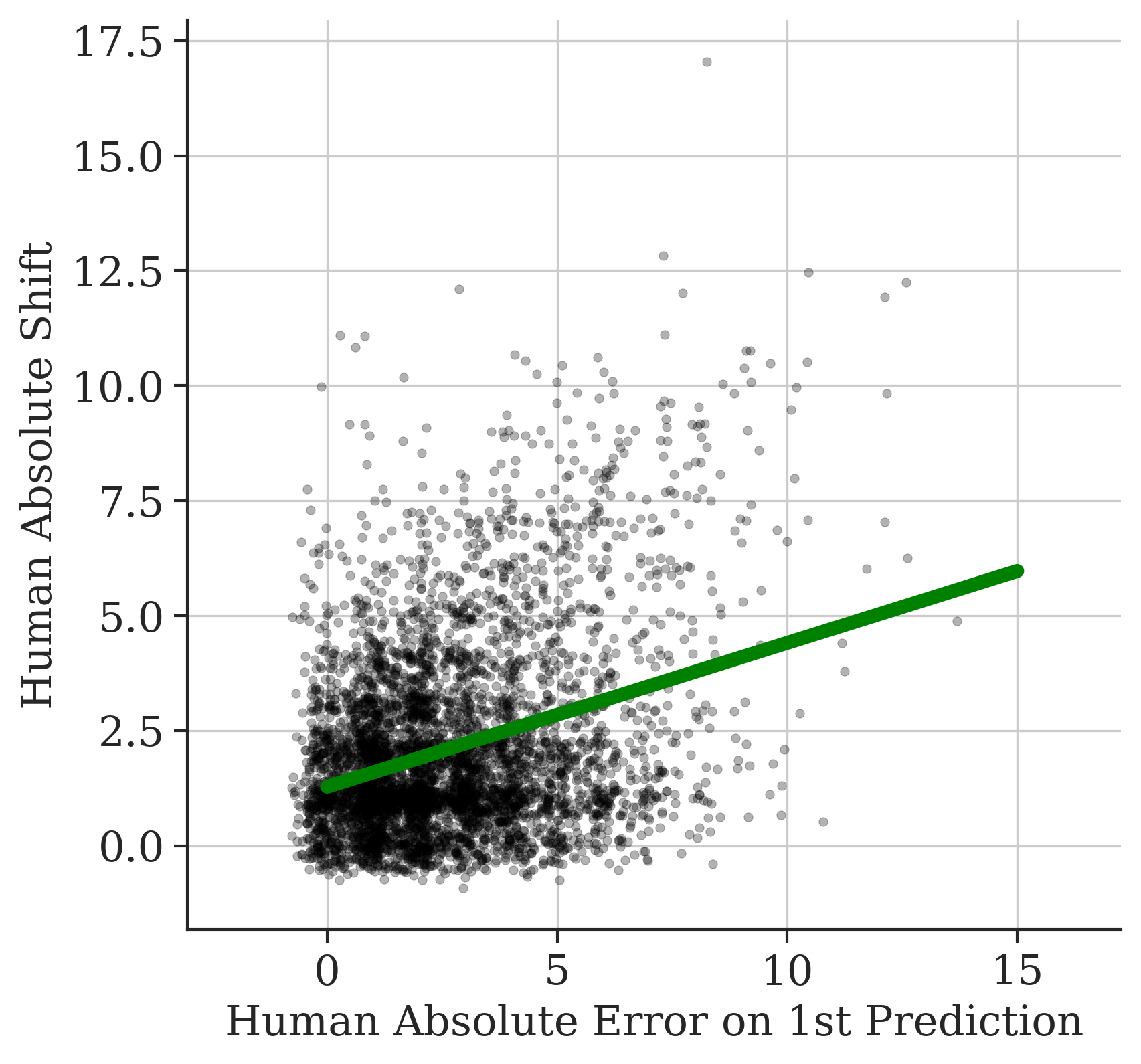}
    \caption{Human absolute response shift versus human absolute error in their first prediction, showing a positive correlation. Participants making larger errors are more heavily influenced by the AI input, leading to a larger shift between their first and second prediction responses ($r=0.31, p<0.001$) -- Gaussian noise ($\mu=0, \sigma=0.3$) added as jitter to scatter plot for better visibility.}
    \label{fig:shift-vs-error}
    \end{center}
\end{figure}

Looking at the same breakdown when humans are paired with a Poor AI, we see that a Good Explanation still results in a significantly better performance by the human compared to having a Poor Explanation ($p<0.01$). However, in this case, the performance with good explanations is not significantly different to that with no explanation at all (see Fig.~\ref{fig:error}), suggesting that the AI quality is more valuable in improving human outcomes, than explanations.

Aside from the effect of our different experiment conditions on human performance in the task, we are also interested in examining how much humans are influenced by the AI's input, given the different conditions. Note that a large influence will not necessarily result in a reduced error, therefore, here we look at the absolute shift in human-predicted values between the first and second responses. Similarly to the error plots, here we look at mean absolute shift across participants for given stimuli, and then consider the grand average over all stimuli, see Fig.~\ref{fig:shift} for a summary.

We see that the Good AI causes a significantly larger shift than the Poor AI in the humans paired with it (paired t-test, with Bonferroni correction, $p<0.05$). Interestingly, breaking down the results for those paired with the Good AI by explanation quality, we see that No Explanation actually leads to a significantly larger absolute shift than receiving a Good Explanation ($p<0.05$), while other explanation conditions do not result in significant differences. When looking at humans paired with the Poor AI, we see that Good Explanations and No Explanation, result in significantly higher response shifts than facing a Poor Explanation ($p<0.0001$), while the inclusion of Good Explanation or lack thereof makes no significant difference. 

Looking at Fig.~\ref{fig:error} and Fig.~\ref{fig:shift} together, we see that in the case of the Good AI, No Explanation leads to a significantly higher response shift, and a significantly higher error, compared to Good Explanations. This could indicate that the Good Explanation is helping the participant work better with the AI, settling on better shifts that lead to lower errors. When paired with a Poor AI, the higher shifts with the Good or No Explanation compared to Poor Explanations, only lead to significantly lower error with the Good Explanation shift. However, this lower error is not significantly lower than that obtained without any explanation. This combined with the Good AI results, suggests that AI performance has a more important role on Human-AI decision making, than explanation quality. Finally, looking directly at absolute response shift versus absolute error on first predictions, we see a positive correlation ($r=0.31, p<0.001$, see Fig.~\ref{fig:shift-vs-error}), indicating that participants with a higher prediction error, are more impacted by the AI's input, resulting in a larger response shift when they submit their 2nd prediction.

\section{Discussion}
In this study we presented a quantitative trust measure derived through a two-step human decision making protocol where humans are exposed to different forms of AI recommendations and explanations. The protocol was designed to be suitable for high-throughput large-scale human-in-the-Loop studies and demonstrated to work efficiently on the Amazon Mechanical Turk with hundreds of users completing the hour long study.

We  manipulated the quality of the AI's predictions exposing users to either 1. a ``Poor AI'' system that had a prediction error higher or at best comparable to that of must human users or 2. a ``Good AI'' system that had  systematically lower prediction error than human users. The underlying AI system was in both cases a deep learner with manipulated learning rates. 

Due to our human subject protocol (see Fig.~\ref{fig:protocol}), users become aware of the AI system qualities through experience during the training phase. In the testing phase we found that human subjects are in general sensitive to the quality of the AI, with the better AI resulting in lower final response errors (see Fig.~\ref{fig:error}) and larger response shifts (see Fig.~\ref{fig:shift}). Furthermore, the amount of impact that the AI prediction quality had on users was bigger the worse subjects performed at the prediction task themselves (see Fig.~\ref{fig:shift-vs-error}). These results are thus consistent with the idea that humans place their trust based on evidence of expertise: better AI prediction ability is more trusted by users, especially if they themselves are not performing as well.

The focus of our thinking was to understand how explanations affect the impact an AI has on the human user. In this initial study we limited ourselves to simple manipulation of this explanation condition: either providing no explanation or one specific form of explanation, here the most important features contributing to the AI's decision through LIME \citep{ribeiro2016should}. In addition, we varied the quality of explanation provided. Different explanation qualities were implemented by confronting users with an explanation generated by the AI system that matched the prediction response (e.g. Good AI prediction and Good AI explanation shown) or one that was mismatched (e.g. Poor AI prediction and Good AI explanation shown -- referred to as a Poor Explanation).

While for this initial study the variations on task, AI operation and explanation are limited by design, these few options already resulted in a combinatorial number of 12 experimental branches (see Fig.~\ref{fig:branches}), that we evaluated and ran on hundreds of users to survey whether these factors are actually contributors to trust in human-AI interactions. This study design allowed us to discover that humans paired with AI input performed best when exposed to the Good AI with Good Explanations (as expected) while those receiving no or poor quality explanations were second best (see Fig.~\ref{fig:error}). Thus, AI prediction quality mattered considerably. Users exposed to combinations of Poor AI or Poor Explanations or both made respectively ever larger errors. We found that users generally trusted the Good AI predictions more (i.e. had a larger response shift, see Fig.~\ref{fig:shift}), and here the presence of an explanation -- irrespective of its quality (!) -- was a bonus on the response shift. This 
suggests that the AI is treated by the subjects as a trustworthy expert based on their experience of its previous prediction (past performance predicting future performance and thus engendering trust). 

In contrast the presence of AI explanations provide a trust bonus (vs. no explanation, see Fig.~\ref{fig:shift}), however, the fact that non-matching explanations also provided an indistinguishable trust bonus from matching ones, suggests that users where reassured by the presence of an explanation but did not evaluate its quality specifically. Conversely, users exposed to the Poor AI system were more sensitive to the quality of the explanations. Thus, in this first study we find that good predictions provide the largest trust bonus, and then  the presence of explanations (but not necessarily their quality) induces larger response shifts. This illustrates how the experience of good predictions and the semblance of explanation may sway users (just as in human-human settings) often more than the details of the explanation. There are many factors that need to be investigated before we can start deducing general design principles, e.g. how sensitive are our results to the presentation or nature of the task (e.g. predictions based on tabular data vs predictions based on images), to the nature of the human decision maker (e.g. common sense type predictions vs expert predictions in safety-critical settings), the nature of the explanations needing to go beyond feature importance, the importance of exposing users  to training vs no training experience with the AI (which depends ultimately on the deployment scenario of an AI recommender sytem). Our preliminary findings for this limited study pose interesting challenges and implications for the field of trustworthy and explainable AI. 

Extensive research has been conducted to examine the factors that influence a human’s trust in autonomous agents 
\citep{parasuraman1997humans,robinette2016overtrust} - it is beyond the scope of this technical focused study to list or enumerate these all, but broadly we can identify machine-related factors, human-related factors and environment-related factors. Our protocol focused on the machine-related factors as the driving force and environment-related factors by giving subjects training experience to get to know the AI, modulating the nature of its presentation of information and its quality. Some human related factors have been captured in terms of demographics and questionnaires for the users who participated in our study but we have excluded them from results as it would be beyond the scope of this short technical paper. Other human factors such as an individual’s overall tendency to trust, so called situational trust, are partially captured by our design, somewhat reflected in our finding that the presence of explanations is influential (but less so than the quality of explanations matter). However how the cost/benefit of the human's final decision influence the degree of trust have not been captured in the presented design. Our protocol contains an explicit training phase, thus the notion of dynamically learned trust (trust formed during an interaction) was measured in steady state, and the transients were not measured. Further demographic analysis of the users participating in our study will reveal how much their ``a priori'' trust in AI  systems, attitude to technology, education and other factors may contribute to a better understanding on how to dynamically personalise AI systems better. 

Crucially our method provides a directly measurable quantitative signal in form of the ``Response Shift'', that we can use  to objectively evaluate and compare different AI explanatory methods on a common currency and task, which will help shape and understand which algorithms will have the most impact on human recommendations and help settle and diversify the debate about which of the logic-based, neurosymbolic, causal and other explanatory approaches are best suited for which applications, tasks and users. Importantly, we obtain a numerical signal that we can feed back directly into a machine learning system to automatically improve and personalise the nature of interaction with the human and optimise trust -- thus our method enables gradient-based learning of trust. Of course, there is nothing that limits our protocol to just measure of trust in AI systems, it could also be used to measure human-human trust interactions objectively using the same protocol.

Our measure of trust is based on a very specific measurement procotol, and it is known that the task design (when is what information provided and in what order) does matter. Here, we chose this specific design as it established a measure of pre-AI baseline on each trial, that we can then differentiate directly with the response of the human post-AI. Building an AI system that engenders trust has many benefits, such as the personalisation of trust to individual ways of reasoning and thinking about trust. However, this in itself bears risks that machine learning researchers need to be aware of, most importantly the risk of users granting blind trust to a system, especially one that has optimised itself to influence the user's trust, where the trust afforded outpaces that actual trustworthiness. Perhaps a lessen that we can take away from this study already is that AI systems and trust in them, are ultimately problems founded in human cognition and psychology, not just AI, as we find that human users can misplace trust in our AI recommendations that are not deserved, or some (a few) fail to trust those that are trustworthy or at least more precise than the human user.

\section*{Acknowledgements}
AS and AAF acknowledge financial support by the EPSRC Network+ \emph{Human-Like Computing} Kick-start grant. VD and HK were both supported in their MSc studies, which formed part of this study, by Google DeepMind MSc scholarships. AAF acknowledges a UKRI Turing AI Fellowship Grant EP/V025449/1 (AAF). The funders were not involved in the design or publication of this study. The authors declare no competing financial interests.


\bibliography{references}

\begin{thebibliography}{40}
\providecommand{\natexlab}[1]{#1}
\providecommand{\url}[1]{\texttt{#1}}
\expandafter\ifx\csname urlstyle\endcsname\relax
  \providecommand{\doi}[1]{doi: #1}\else
  \providecommand{\doi}{doi: \begingroup \urlstyle{rm}\Url}\fi

\bibitem[Ajenaghughrure et~al.(2020)Ajenaghughrure, Sousa, and
  Lamas]{ajenaghughrure2020measuring}
Ajenaghughrure, I.~B., Sousa, S. D.~C., and Lamas, D.
\newblock Measuring trust with psychophysiological signals: A systematic
  mapping study of approaches used.
\newblock \emph{Multimodal Technologies and Interaction}, 4\penalty0
  (3):\penalty0 63, 2020.

\bibitem[Akash et~al.(2018)Akash, Hu, Jain, and Reid]{akash2018classification}
Akash, K., Hu, W.-L., Jain, N., and Reid, T.
\newblock A classification model for sensing human trust in machines using eeg
  and gsr.
\newblock \emph{ACM Transactions on Interactive Intelligent Systems (TiiS)},
  8\penalty0 (4):\penalty0 1--20, 2018.

\bibitem[Alam \& Mueller(2021)Alam and Mueller]{alam2021examining}
Alam, L. and Mueller, S.
\newblock Examining the effect of explanation on satisfaction and trust in ai
  diagnostic systems.
\newblock \emph{BMC medical informatics and decision making}, 21\penalty0
  (1):\penalty0 1--15, 2021.

\bibitem[Alarcon et~al.(2021)Alarcon, Gibson, Jessup, and
  Capiola]{alarcon2021exploring}
Alarcon, G.~M., Gibson, A.~M., Jessup, S.~A., and Capiola, A.
\newblock Exploring the differential effects of trust violations in human-human
  and human-robot interactions.
\newblock \emph{Applied Ergonomics}, 93:\penalty0 103350, 2021.

\bibitem[Anton et~al.(2022)Anton, Oesterreich, Fitte, and
  Teuteberg]{anton2022trust}
Anton, E., Oesterreich, T.~D., Fitte, C., and Teuteberg, F.
\newblock Trust recipes for enhancing the intention to adopt conversational
  agents for disease diagnosis: An fsqca approach.
\newblock In \emph{Proceedings of the 55th Hawaii International Conference on
  System Sciences}, 2022.

\bibitem[Bansal et~al.(2021)Bansal, Wu, Zhou, Fok, Nushi, Kamar, Ribeiro, and
  Weld]{bansal2021does}
Bansal, G., Wu, T., Zhou, J., Fok, R., Nushi, B., Kamar, E., Ribeiro, M.~T.,
  and Weld, D.
\newblock Does the whole exceed its parts? the effect of ai explanations on
  complementary team performance.
\newblock In \emph{Proceedings of the 2021 CHI Conference on Human Factors in
  Computing Systems}, pp.\  1--16, 2021.

\bibitem[Bedoya et~al.(2019)Bedoya, Clement, Phelan, Steorts, O’Brien, and
  Goldstein]{bedoya2019minimal}
Bedoya, A.~D., Clement, M.~E., Phelan, M., Steorts, R.~C., O’Brien, C., and
  Goldstein, B.~A.
\newblock Minimal impact of implemented early warning score and best practice
  alert for patient deterioration.
\newblock \emph{Critical care medicine}, 47\penalty0 (1):\penalty0 49, 2019.

\bibitem[Beierholm et~al.(2008)Beierholm, Shams, Ma, and
  Koerding]{beierholm2008comparing}
Beierholm, U., Shams, L., Ma, W.~J., and Koerding, K.
\newblock Comparing bayesian models for multisensory cue combination without
  mandatory integration.
\newblock In \emph{Advances in neural information processing systems}, pp.\
  81--88, 2008.

\bibitem[Bu{\c{c}}inca et~al.(2021)Bu{\c{c}}inca, Malaya, and
  Gajos]{buccinca2021trust}
Bu{\c{c}}inca, Z., Malaya, M.~B., and Gajos, K.~Z.
\newblock To trust or to think: cognitive forcing functions can reduce
  overreliance on ai in ai-assisted decision-making.
\newblock \emph{Proceedings of the ACM on Human-Computer Interaction},
  5\penalty0 (CSCW1):\penalty0 1--21, 2021.

\bibitem[Cortez \& Silva(2008)Cortez and Silva]{cortez2008using}
Cortez, P. and Silva, A. M.~G.
\newblock Using data mining to predict secondary school student performance.
\newblock In \emph{Proceedings of 5th Annual Future Business Technology
  Conference, Porto, 2008}. EUROSIS-ETI, 2008.

\bibitem[{de Leeuw JR}(2015)]{deLeeuwJR2015JsPsych:Browser}
{de Leeuw JR}.
\newblock {jsPsych: A JavaScript library for creating behavioral experiments in
  a Web browser}.
\newblock \emph{Behavior Research Methods}, 47\penalty0 (1):\penalty0 1--12,
  2015.
\newblock ISSN 15543528.
\newblock \doi{10.3758/s13428-014-0458-y}.
\newblock URL \url{https://pubmed.ncbi.nlm.nih.gov/24683129/}.

\bibitem[Dietvorst et~al.(2015)Dietvorst, Simmons, and
  Massey]{dietvorst2015algorithm}
Dietvorst, B.~J., Simmons, J.~P., and Massey, C.
\newblock Algorithm aversion: People erroneously avoid algorithms after seeing
  them err.
\newblock \emph{Journal of Experimental Psychology: General}, 144\penalty0
  (1):\penalty0 114, 2015.

\bibitem[Doshi-Velez \& Kim(2017)Doshi-Velez and Kim]{doshi2017towards}
Doshi-Velez, F. and Kim, B.
\newblock Towards a rigorous science of interpretable machine learning.
\newblock \emph{arXiv preprint arXiv:1702.08608}, 2017.

\bibitem[Drnec et~al.(2016)Drnec, Marathe, Lukos, and Metcalfe]{drnec2016trust}
Drnec, K., Marathe, A.~R., Lukos, J.~R., and Metcalfe, J.~S.
\newblock From trust in automation to decision neuroscience: applying cognitive
  neuroscience methods to understand and improve interaction decisions involved
  in human automation interaction.
\newblock \emph{Frontiers in human neuroscience}, 10:\penalty0 290, 2016.

\bibitem[Faisal et~al.(2008)Faisal, Selen, and Wolpert]{faisal2008noise}
Faisal, A.~A., Selen, L.~P., and Wolpert, D.~M.
\newblock Noise in the nervous system.
\newblock \emph{Nature reviews neuroscience}, 9\penalty0 (4):\penalty0
  292--303, 2008.

\bibitem[Gille et~al.(2020)Gille, Jobin, and Ienca]{gille2020we}
Gille, F., Jobin, A., and Ienca, M.
\newblock What we talk about when we talk about trust: Theory of trust for ai
  in healthcare.
\newblock \emph{Intelligence-Based Medicine}, 1:\penalty0 100001, 2020.

\bibitem[Gupta et~al.(2020)Gupta, Hajika, Pai, Duenser, Lochner, and
  Billinghurst]{gupta2020measuring}
Gupta, K., Hajika, R., Pai, Y.~S., Duenser, A., Lochner, M., and Billinghurst,
  M.
\newblock Measuring human trust in a virtual assistant using physiological
  sensing in virtual reality.
\newblock In \emph{2020 IEEE Conference on Virtual Reality and 3D User
  Interfaces (VR)}, pp.\  756--765. IEEE, 2020.

\bibitem[Jones(2016)]{jones2016tutorial}
Jones, P.~R.
\newblock A tutorial on cue combination and signal detection theory: Using
  changes in sensitivity to evaluate how observers integrate sensory
  information.
\newblock \emph{Journal of Mathematical Psychology}, 73:\penalty0 117--139,
  2016.

\bibitem[Kingma \& Ba(2015)Kingma and Ba]{kingma2014adam}
Kingma, D.~P. and Ba, J.
\newblock Adam: A method for stochastic optimization.
\newblock In \emph{Proceedings of the International Conference on Learning
  Representations}, 2015.

\bibitem[Komorowski et~al.(2018)Komorowski, Celi, Badawi, Gordon, and
  Faisal]{komorowski2018artificial}
Komorowski, M., Celi, L.~A., Badawi, O., Gordon, A.~C., and Faisal, A.~A.
\newblock The artificial intelligence clinician learns optimal treatment
  strategies for sepsis in intensive care.
\newblock \emph{Nature medicine}, 24\penalty0 (11):\penalty0 1716--1720, 2018.

\bibitem[K{\"o}rding \& Wolpert(2006)K{\"o}rding and
  Wolpert]{kording2006bayesian}
K{\"o}rding, K.~P. and Wolpert, D.~M.
\newblock Bayesian decision theory in sensorimotor control.
\newblock \emph{Trends in cognitive sciences}, 10\penalty0 (7):\penalty0
  319--326, 2006.

\bibitem[Lai \& Tan(2019)Lai and Tan]{lai2019human}
Lai, V. and Tan, C.
\newblock On human predictions with explanations and predictions of machine
  learning models: A case study on deception detection.
\newblock In \emph{Proceedings of the conference on fairness, accountability,
  and transparency}, pp.\  29--38, 2019.

\bibitem[Lu \& Sarter(2019)Lu and Sarter]{lu2019eye}
Lu, Y. and Sarter, N.
\newblock Eye tracking: a process-oriented method for inferring trust in
  automation as a function of priming and system reliability.
\newblock \emph{IEEE Transactions on Human-Machine Systems}, 49\penalty0
  (6):\penalty0 560--568, 2019.

\bibitem[Maloney \& Mamassian(2009)Maloney and Mamassian]{maloney2009bayesian}
Maloney, L.~T. and Mamassian, P.
\newblock Bayesian decision theory as a model of human visual perception:
  Testing bayesian transfer.
\newblock \emph{Visual neuroscience}, 26\penalty0 (1):\penalty0 147--155, 2009.

\bibitem[Okamura \& Yamada(2020)Okamura and Yamada]{okamura2020adaptive}
Okamura, K. and Yamada, S.
\newblock Adaptive trust calibration for human-ai collaboration.
\newblock \emph{Plos one}, 15\penalty0 (2):\penalty0 e0229132, 2020.

\bibitem[Parasuraman \& Riley(1997)Parasuraman and
  Riley]{parasuraman1997humans}
Parasuraman, R. and Riley, V.
\newblock Humans and automation: Use, misuse, disuse, abuse.
\newblock \emph{Human factors}, 39\penalty0 (2):\penalty0 230--253, 1997.

\bibitem[Paszke et~al.(2019)Paszke, Gross, Massa, Lerer, Bradbury, Chanan,
  Killeen, Lin, Gimelshein, Antiga, et~al.]{paszke2019pytorch}
Paszke, A., Gross, S., Massa, F., Lerer, A., Bradbury, J., Chanan, G., Killeen,
  T., Lin, Z., Gimelshein, N., Antiga, L., et~al.
\newblock Pytorch: An imperative style, high-performance deep learning library.
\newblock \emph{Advances in neural information processing systems},
  32:\penalty0 8026--8037, 2019.

\bibitem[Paulhus et~al.(2007)Paulhus, Vazire, et~al.]{paulhus2007self}
Paulhus, D.~L., Vazire, S., et~al.
\newblock The self-report method.
\newblock \emph{Handbook of research methods in personality psychology},
  1\penalty0 (2007):\penalty0 224--239, 2007.

\bibitem[Payrovnaziri et~al.(2020)Payrovnaziri, Chen, Rengifo-Moreno, Miller,
  Bian, Chen, Liu, and He]{payrovnaziri2020explainable}
Payrovnaziri, S.~N., Chen, Z., Rengifo-Moreno, P., Miller, T., Bian, J., Chen,
  J.~H., Liu, X., and He, Z.
\newblock Explainable artificial intelligence models using real-world
  electronic health record data: a systematic scoping review.
\newblock \emph{Journal of the American Medical Informatics Association},
  27\penalty0 (7):\penalty0 1173--1185, 2020.

\bibitem[Ribeiro et~al.(2016)Ribeiro, Singh, and Guestrin]{ribeiro2016should}
Ribeiro, M.~T., Singh, S., and Guestrin, C.
\newblock ``why should i trust you?'' explaining the predictions of any
  classifier.
\newblock In \emph{Proceedings of the 22nd ACM SIGKDD international conference
  on knowledge discovery and data mining}, pp.\  1135--1144, 2016.

\bibitem[Riley(2018)]{riley2018operator}
Riley, V.
\newblock Operator reliance on automation: Theory and data.
\newblock In \emph{Automation and human performance: Theory and applications},
  pp.\  19--35. CRC Press, 2018.

\bibitem[Robinette et~al.(2016)Robinette, Li, Allen, Howard, and
  Wagner]{robinette2016overtrust}
Robinette, P., Li, W., Allen, R., Howard, A.~M., and Wagner, A.~R.
\newblock Overtrust of robots in emergency evacuation scenarios.
\newblock In \emph{2016 11th ACM/IEEE International Conference on Human-Robot
  Interaction (HRI)}, pp.\  101--108. IEEE, 2016.

\bibitem[Rudin(2019)]{rudin2019stop}
Rudin, C.
\newblock Stop explaining black box machine learning models for high stakes
  decisions and use interpretable models instead.
\newblock \emph{Nature Machine Intelligence}, 1\penalty0 (5):\penalty0
  206--215, 2019.

\bibitem[Tenenbaum et~al.(2006)Tenenbaum, Griffiths, and
  Kemp]{tenenbaum2006theory}
Tenenbaum, J.~B., Griffiths, T.~L., and Kemp, C.
\newblock Theory-based bayesian models of inductive learning and reasoning.
\newblock \emph{Trends in cognitive sciences}, 10\penalty0 (7):\penalty0
  309--318, 2006.

\bibitem[Toreini et~al.(2020)Toreini, Aitken, Coopamootoo, Elliott, Zelaya, and
  Van~Moorsel]{toreini2020relationship}
Toreini, E., Aitken, M., Coopamootoo, K., Elliott, K., Zelaya, C.~G., and
  Van~Moorsel, A.
\newblock The relationship between trust in ai and trustworthy machine learning
  technologies.
\newblock In \emph{Proceedings of the 2020 conference on fairness,
  accountability, and transparency}, pp.\  272--283, 2020.

\bibitem[Vollmer et~al.(2018)Vollmer, Mateen, Bohner, Kir{\'a}ly, Ghani,
  Jonsson, Cumbers, Jonas, McAllister, Myles, et~al.]{vollmer2018machine}
Vollmer, S., Mateen, B.~A., Bohner, G., Kir{\'a}ly, F.~J., Ghani, R., Jonsson,
  P., Cumbers, S., Jonas, A., McAllister, K.~S., Myles, P., et~al.
\newblock Machine learning and ai research for patient benefit: 20 critical
  questions on transparency, replicability, ethics and effectiveness.
\newblock \emph{arXiv preprint arXiv:1812.10404}, 2018.

\bibitem[Walker et~al.(2019)Walker, Wang, Martens, and Verwey]{walker2019gaze}
Walker, F., Wang, J., Martens, M.~H., and Verwey, W.~B.
\newblock Gaze behaviour and electrodermal activity: Objective measures of
  drivers' trust in automated vehicles.
\newblock \emph{Transportation research part F: traffic psychology and
  behaviour}, 64:\penalty0 401--412, 2019.

\bibitem[Wang et~al.(2018)Wang, Hussein, Rojas, Shafi, and Abbass]{wang2018eeg}
Wang, M., Hussein, A., Rojas, R.~F., Shafi, K., and Abbass, H.~A.
\newblock Eeg-based neural correlates of trust in human-autonomy interaction.
\newblock In \emph{2018 IEEE Symposium Series on Computational Intelligence
  (SSCI)}, pp.\  350--357. IEEE, 2018.

\bibitem[Wiens et~al.(2019)Wiens, Saria, Sendak, Ghassemi, Liu, Doshi-Velez,
  Jung, Heller, Kale, Saeed, et~al.]{wiens2019no}
Wiens, J., Saria, S., Sendak, M., Ghassemi, M., Liu, V.~X., Doshi-Velez, F.,
  Jung, K., Heller, K., Kale, D., Saeed, M., et~al.
\newblock Do no harm: a roadmap for responsible machine learning for health
  care.
\newblock \emph{Nature medicine}, 25\penalty0 (9):\penalty0 1337--1340, 2019.

\bibitem[Yuille \& B{\"u}lthoff(1996)Yuille and
  B{\"u}lthoff]{yuille1996bayesian}
Yuille, A. and B{\"u}lthoff, H.
\newblock Bayesian decision theory and psychophysics.
\newblock In \emph{Perception as Bayesian Inference}, pp.\  123--161. Cambridge
  University Press, 1996.

\end{thebibliography}
\bibliographystyle{bibstyle}

\newpage
\appendix
\onecolumn
\setcounter{figure}{0}
\makeatletter 
\renewcommand{\thefigure}{S\@arabic\c@figure}
\makeatother
\section{Overview of grade prediction dataset features}
\label{appendix:tabular_features}

\begin{table}[h]
\label{tab:tabular_features}
\caption{Tabular features, their category in the web experiment, and their possible values. \emph{Note}: Based on table 1 in \citep{cortez2008using}. Values column represents values as they were shown to participants in the experiment, not the values in the dataset. Most nominal values were encoded with integers in the dataset.}
\centering
\begin{tabular}{@{}lll@{}}
\toprule
\textbf{Feature name}            & \textbf{Category} & \textbf{Values}                                                               \\ \midrule
Guardian                         & Family            & Mother, father, other                                                         \\
Mother’s job                     & Family            & At home, civil services, teacher, healthcare, other                           \\
Father’s job                     & Family            & At home, civil services, teacher, healthcare, other                           \\
Parent’s cohabitation status     & Family            & Together, apart                                                               \\
Mother’s education level         & Family            & None, up to 4th/9th grade, secondary ed., higher ed. \\
Father’s education level         & Family            & None, up to 4th/9th grade, secondary ed., higher ed. \\
Family size                      & Family            & Three or fewer, more than three                                               \\
Family educational support       & Family            & Yes, no                                                                       \\
Quality of family relationships  & Family            & Very bad, bad, average, good, excellent                                       \\
School name                      & School            & GP (Gabriel Pereira), MS (Mousinho da Silveira)                               \\
Reason to choose this school     & School            & Closeness, school reputation, course prefs., other                   \\
Number of absences               & School            & 0 - 93                                                                        \\
Travel time to school            & School            & $<$15min, 15-30min, 0.5-1h, $>$1h            \\
Weekly study time                & School            & $<$2h, 2-5h, 5-10h, $>$10h            \\
Past class failures              & School            & 0 – 4                                                                         \\
Extra educational school support & School            & Yes, no                                                                       \\
Wants to pursue higher education & School            & Yes, no                                                                       \\
Extra paid classes               & School            & Yes, no                                                                       \\
Attended nursery school          & School            & Yes, no                                                                       \\
Age                              & Other             & 15 – 22                                                                       \\
Gender                           & Other             & Female, male                                                                  \\
Address                          & Other             & Rural, urban                                                                  \\
In a romantic relationship       & Other             & Yes, no                                                                       \\
Current health status            & Other             & Very bad, bad, average,   good, excellent                                     \\
Extra-curricular activities      & Other             & Yes, no                                                                       \\
Internet access at home          & Other             & Yes, no                                                                       \\
Free time after school           & Other             & Very low, low, average, high, very high                                       \\
Going out with friends           & Other             & Almost never, not often, sometimes, often, v. often                         \\
Workday alcohol consumption      & Other             & Very low, low, average, high, very high                                       \\
Weekend alcohol consumption      & Other             & Very low, low, average, high, very high                                       \\ \bottomrule
\end{tabular}
\end{table}
\pagebreak
\section{Experiment interface}
\label{appendix:experiment_interface}

\begin{figure}[h]
    \begin{center}
    \includegraphics[width=0.7\textwidth]{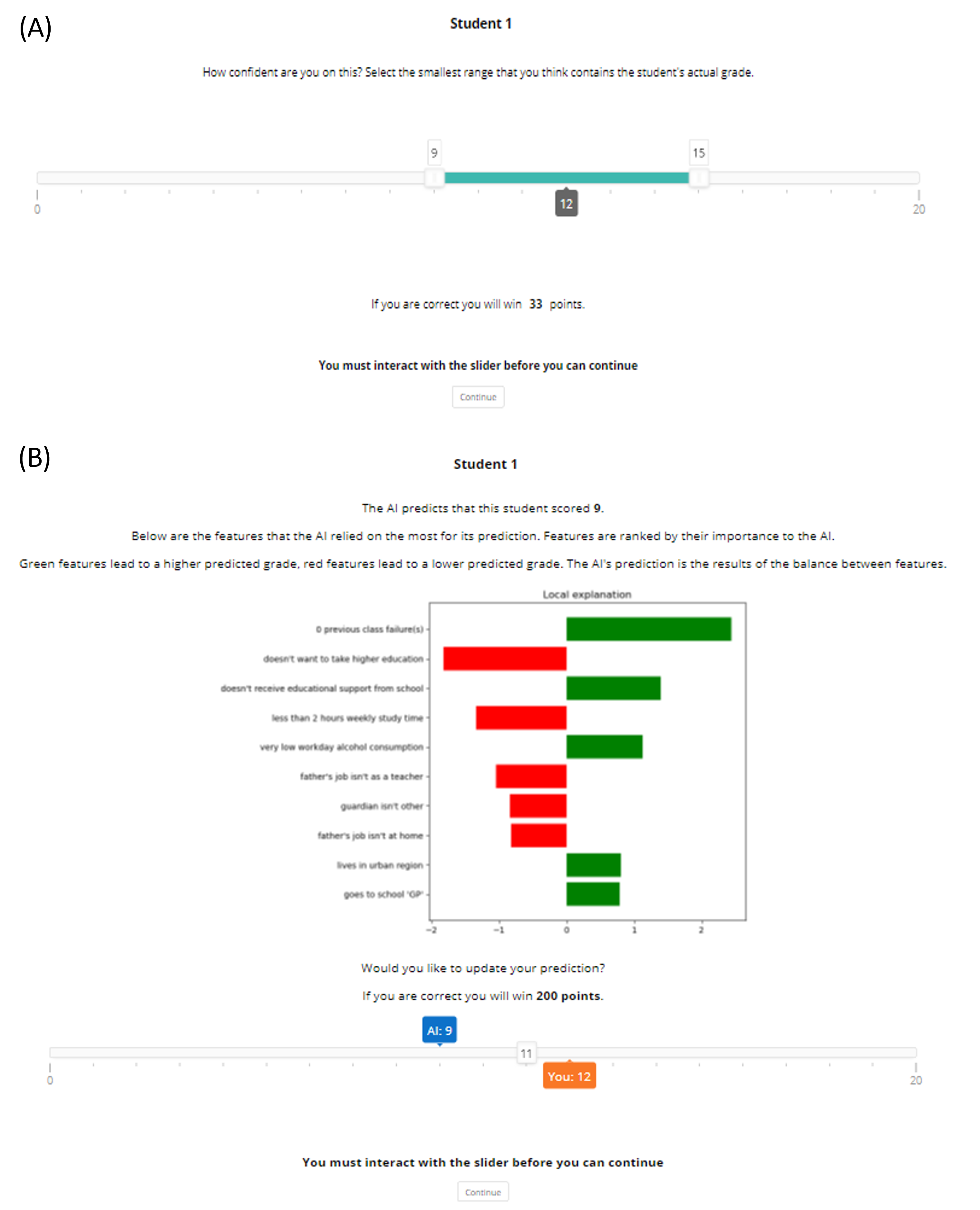}
    \caption{Further examples of our web-based experiment interface. (A) Slider for participants to indicate the confidence range for their prediction, note that the interface shows their expected score if they are correct, and how it reduces with a larger range. (B) AI prediction and explanation provided to participants. LIME explanation is shown, with the AI's prediction shown with the participant's first prediction on the grade bar. Participants are then allowed to respond with a second prediction.}
    \label{fig:further-interface}
    \end{center}
\end{figure}

\section{Questionnaire on attitude towards AI}
\label{appendix:questionnaire}

The questions that all participants are asked to answer are:

\begin{itemize}
    \item ``How do you feel towards Artificial Intelligence (AI) in general?''
    \item ``How do you feel about AI being used to help make decisions in medical settings?''
    \item ``How do you feel about AI being used to help make decisions in education settings?''
\end{itemize}
\pagebreak
\section{Demographics}
\label{appendix:demographics}

\begin{figure}[h]
    \begin{center}
    \includegraphics[width=0.50\textwidth]{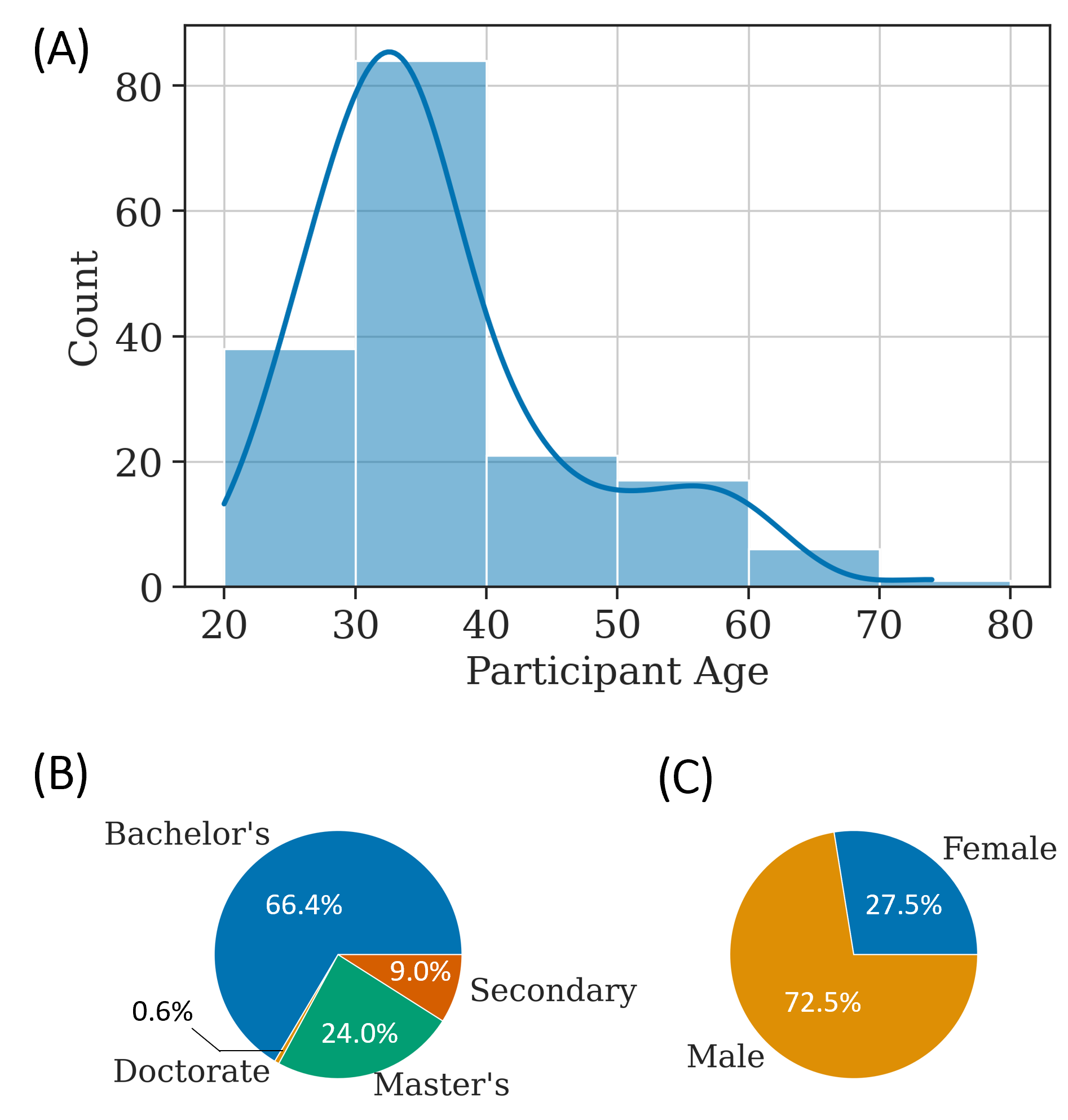}
    \caption{The demographics of our 167 participants.}
    \label{fig:demographics}
    \end{center}
\end{figure}

\end{document}